%% file: ms.tex
\documentclass{emulateapj}
\usepackage{natbib}
\usepackage{apjfonts}
\usepackage{multirow}

\newcommand{\rl}{$R_{\rm BLR} - L$}

\newcommand{\msun}{M$_{\odot}$}
\newcommand{\sersic}{S\'{e}rsic}

\newcommand{\hst}{{\it HST}}
\newcommand{\flam}{erg\,s$^{-1}$\,cm$^{-2}$\,\AA$^{-1}$}

\shorttitle{The Host-Galaxy of UGC\,06728}
\shortauthors{Bentz et al.}

\received{}
\accepted{}

\begin{document}

\title{The Host Galaxy of the Dwarf Seyfert UGC\,06728}

\author{ Misty~C.~Bentz\altaffilmark{1}
}

\altaffiltext{1}{Department of Physics and Astronomy,
		 Georgia State University,
		 Atlanta, GA 30303, USA;
		 bentz@astro.gsu.edu}

\begin{abstract}

We present multi-color high-resolution imaging of the host galaxy of the dwarf Seyfert UGC\,06728.  As the lowest-mass black hole to be described with both a direct mass constraint and a spin constraint, UGC\,06728 is an important source for comparison with black hole evolutionary models, yet little is known about the host galaxy.  Using {\it Hubble Space Telescope} imaging in the optical and near-infrared, we find that UGC\,06728 is a barred lenticular (SB0) galaxy with prominent ansae at the ends of the bar.  We cleanly separated the AGN from the resolved galaxy with two-dimensional image decompositions, thus allowing accurate surface brightness profiles to be derived in all filters from the outer edge of the galaxy all the way into the nucleus.  Based on a sample of 51 globular cluster candidates identified in the images, the globular cluster luminosity function predicts a distance to UCG\,06728 of $32.5\pm3.5$\,Mpc. Combining the galaxy photometry with the distance estimate, we derive a starlight-corrected AGN luminosity, the absolute magnitude of the galaxy, and a constraint on the galaxy stellar mass of $\log M_{\star}/M_{\odot}=9.9\pm0.2$.

\end{abstract}


\section{Introduction}

It is a truth universally acknowledged that a massive galaxy must not be in want of a supermassive black hole.  But when it comes to low-mass and dwarf galaxies, the situation is rather less clear.

UGC\,06728 is a relatively unstudied dwarf Seyfert~1.  At a redshift of only $z=0.0065$, it is one of the nearest broad-lined AGNs.  As part of an ongoing effort to expand the sample of galaxies with black hole masses determined from both reverberation mapping and stellar dynamical modeling, UGC\,06728 was targeted for reverberation mapping with the Apache Point Observatory (APO) 3.5-m telescope in Spring 2015.  Spectrophotometric monitoring was carried out over the course of 45 days, resulting in a measured time delay between the broad H$\beta$ emission line and the continuum of $1.4\pm0.8$\,days and a corresponding black hole mass of $M_{\rm BH}=(7.1\pm4.0) \times 10^5$\,\msun\ \citep{bentz16b}.  

As part of the reverberation mapping campaign, photometric monitoring was also carried out on the APO 3.5-m telescope and at Georgia State University's Hard Labor Creek Observatory.  Stacked images from the photometric monitoring hinted at a moderately-inclined disk galaxy with a stellar mass $<10^{10}$\,\msun\ based on the $g-r$ color and initial attempts to separate the galaxy light from the AGN.  However, the poor spatial resolution of the seeing-limited images, the small angular diameter of the galaxy, and the blending of the galaxy with the central AGN limited the information that could be constrained about the host. 

Interestingly, while the estimated bulge-to-total ratio for UGC\,06728 suggested a late-type galaxy \citep{bentz16b}, in 1\,hour of exposure time with the 100-m Green Bank Telescope, there was no detection of HI 21\,cm emission from the galaxy  \citep{robinson19}.  There also appears to be a deficit of dust, given the moderate inclination of the galaxy and the detection of no obscuration along the line of sight to the nucleus in Suzaku observations \citep{walton13}. 

As the smallest of the supermassive black holes with {\bf both} a direct mass constraint and a spin constraint ($a>0.7; $\citealt{walton13}), UGC\,06728 is an intriguing nearby galaxy for testing black hole evolutionary models.  For example, \citet{volonteri13} predict that low-mass black holes in local, gas-rich galaxies should be spinning slowly, contrary to what is observed for UGC\,06728.  On the other hand, \citet{sesana14} predict that disk instabilities should spin up local, low-mass black holes.

Given the difficulties in accurately characterizing the host galaxy using ground-based images, we undertook a multi-color imaging program with the {\it Hubble Space Telescope} (\hst) with a variety of goals.  The first was to securely determine the morphology of the galaxy so that the properties of the central black hole could be better interpreted in light of evolutionary models.  We also planned our observations to constrain the central stellar surface brightness, the stellar mass-to-light ratio of the galaxy through commonly used colors, and to possibly estimate the distance to the galaxy using its associated globular clusters.  We describe our findings below.

\begin{figure*}
\plotone{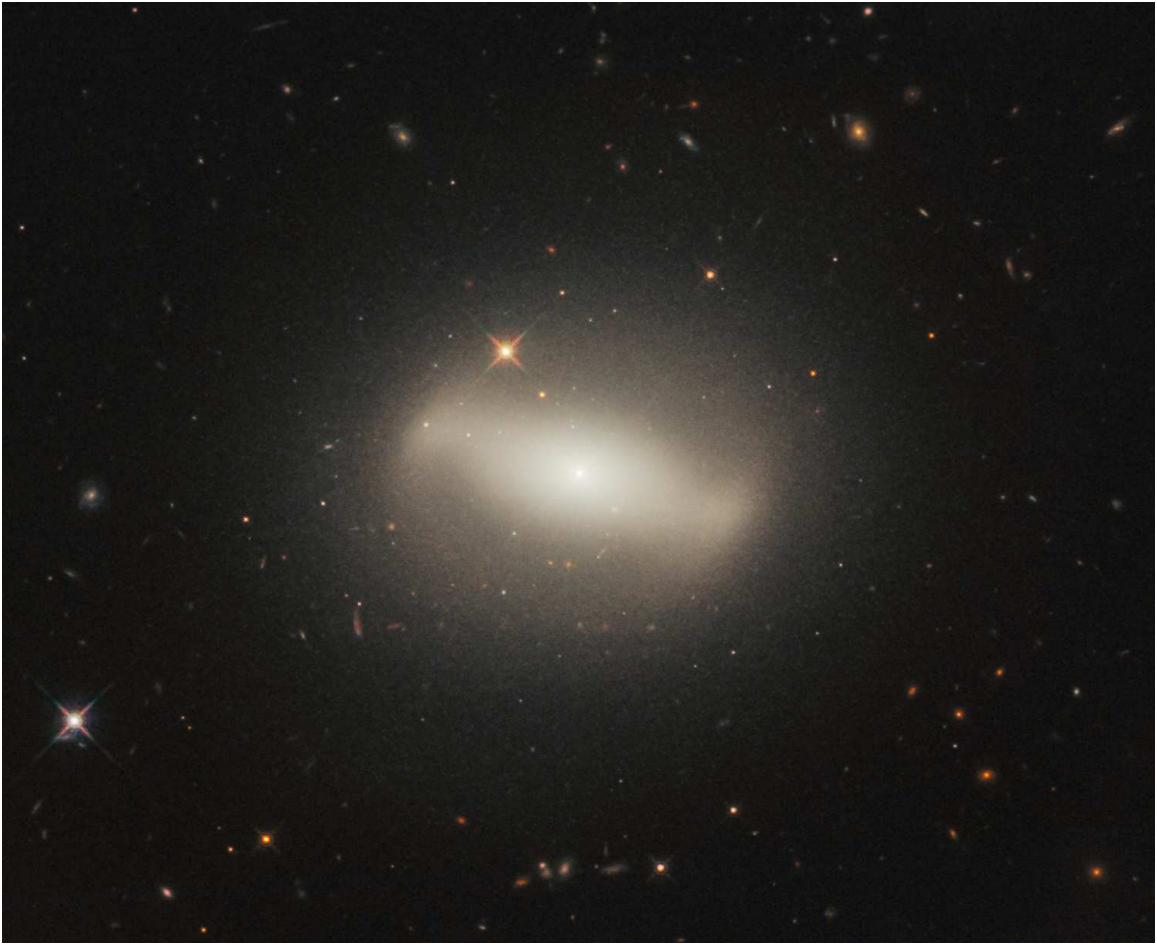}
\caption{Composite image of UGC\,06728 created by combining the F547M, F814W, and F160W images.  The image size is $2\farcm0 \times 1\farcm6$ and is oriented with North rotated $45.27^{\circ}$ clockwise from the top.  Image credit: Judy Schmidt. }
\label{fig:pretty}
\end{figure*}

\section{Observations}

UGC\,06728 is located in the direction of Camelopardalis at $\alpha=$11:45:16.0, $\delta=+$79:40:53, $z=0.00652$. Observations were conducted with the {\it Hubble Space Telescope} during Cycle 25.  Two sequential orbits on UT date 2018 March 16 were dedicated to Wide Field Camera 3 (WFC3) UVIS and IR imaging through the F547M, F814W, and F160W filters.  A summary of the observations is given in Table~\ref{tab:obs}.

\begin{deluxetable}{lccc}
\input{table.obs.tex}

\end{deluxetable}

The F547M and F814W images were executed with 2-point dither patterns to facilitate the rejection of cosmic rays and detector artifacts.  At each point in the pattern, a set of short and long exposures was acquired.  The short exposures allow the bright AGN PSF to be accurately characterized without the effects of saturation, while the long exposures probe the fainter, extended host-galaxy light.  For the F547M images, the adopted exposure times were 75.0\,s and 810.0\,s, and for the F814W images, the adopted exposure times were 75.0\,s and 900.0\,s.  Small post flashes were applied to each exposure to ensure that the background levels were sufficient to promote good charge transfer efficiency during readout.

Four exposures through the F160W filter were obtained using the STEP50 exposure ramp with 14 samples and resulting in exposure times of 399.2\,s.  The exposures were executed with a four-point dither pattern to improve the PSF sampling.  

The UVIS detectors with the F547M and F814W filters provide images with a field of view of $2\farcm6 \times 2\farcm6$ and pixel scale of $0\farcs04$.  Given the compact nature of the galaxy as seen in ground-based imaging, the galaxy was centered on the top UVIS1 detector, and a loose ORIENT constraint was imposed to allow the major axis of the galaxy to roughly follow the long axis of the detector.  The F160W images were centered on the IR detector and resulted in images with a field of view of $2\farcm05 \times 2\farcm27$ and a pixel scale of $0\farcs128$.  

The \hst\ pipeline carried out the basic image reductions, including bias removal, flat fielding, cosmic ray removal, and post-flash removal.   Working on the reduced ({\tt\_flt}) images, we corrected the long UVIS exposures for saturation in the AGN PSF by making use of the linear response of the detector.  For each long exposure, the saturated pixels in the nucleus were clipped out and replaced by the same pixels from the corresponding short exposure, after scaling up the counts by the exposure time ratio.  After correction for saturation, all of the exposures taken through the same filter were distortion-corrected, drizzled, and combined into a single image with {\sc AstroDrizzle}.  In these final, combined images, the F814W and F160W images were drizzled to match the reference frame of the F547M images to aid in further analysis.

Based on the final images, the morphology of the galaxy is revealed to be a barred lenticular (SB0; Figure~\ref{fig:pretty}).

\begin{deluxetable*}{clcccccccccr}
\input{table.galfit.tex}
\tablecomments{Reported magnitudes are based on the following adopted vegamag zeropoints: $zpt_{F547M}=24.764$\,mag, $zpt_{F814W}=24.684$\,mag, $zpt_{F160W}=24.690$\,mag.  Values in square brackets were held fixed during the fitting process. The fraction of the total galaxy flux contributed by each morphological component is given in the column $f/f_{\rm tot}$.}
\end{deluxetable*}

\begin{figure*}[ht!]
\epsscale{1.15}
\plotone{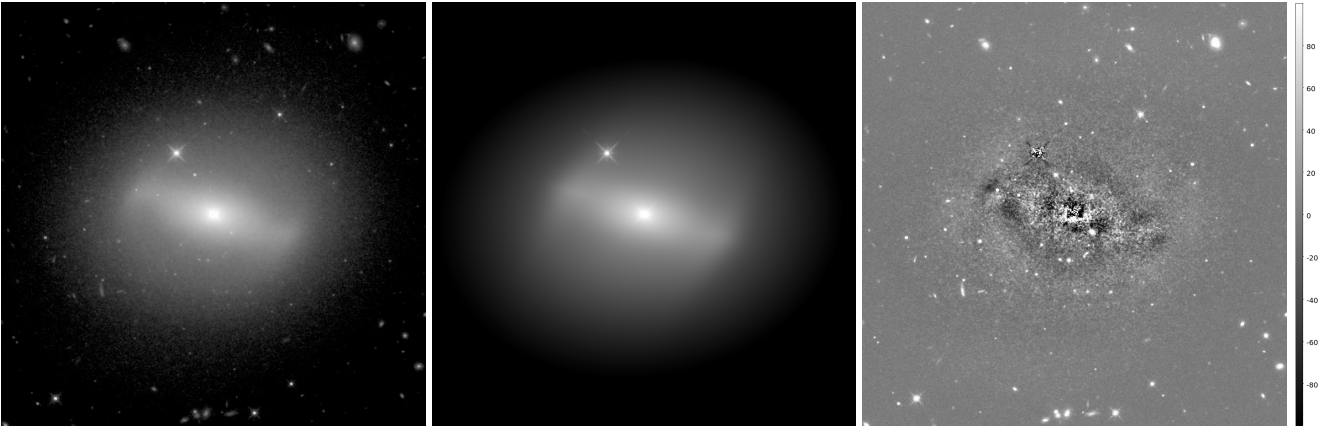}
\caption{F160W image ({\it left}), best-fit model ({\it center}), and residuals after subtracting the model from the image ({\it right}).  The image and model are both displayed with a logarithmic stretch.  The residuals are displayed with a linear stretch centered around zero counts and show regions of oversubtraction and undersubtraction according to the scale bar on the far right.  Each panel is 88\arcsec\ $\times$ 88\arcsec.}
\label{fig:galfit}
\end{figure*}

\section{Image Decompositions}

Two-dimensional decompositions of the galaxy images were carried out with {\sc galfit} \citep{peng02,peng10}.  For each image, the galaxy was fit with several general \citet{sersic68} functions to represent the different morphological components of the galaxy.  An exponential disk profile is a special case of the \sersic\ function where the index has a value of $n=1.0$, while galaxy bulges generally have $n>1$ and bars generally have $n<1$.  The point spread function (PSF) of each image was modeled with {\sc StarFit} \citep{hamilton14} on the individual frames and then drizzled together in the same way as the science images. {\sc StarFit} is based on {\sc TinyTim} \citet{krist93} but attempts to include the effects of spacecraft breathing and focus changes by adjusting the PSF to match a designated field star.  

The prominent ansae at the ends of the bar in UGC\,06728 caused difficulties with the initial image fitting, so an additional \sersic\ component with a high ellipticity was included, with inner and outer truncation functions rotated relative to each other to remove most of the light while leaving two patches to simulate the bright "handles" at the ends of the bar.  Additionally, we found that it was necessary to include the disky/boxy parameter to improve the model fits and the image residuals, especially in the nucleus of the galaxy.

\begin{figure*}[t!]
\plotone{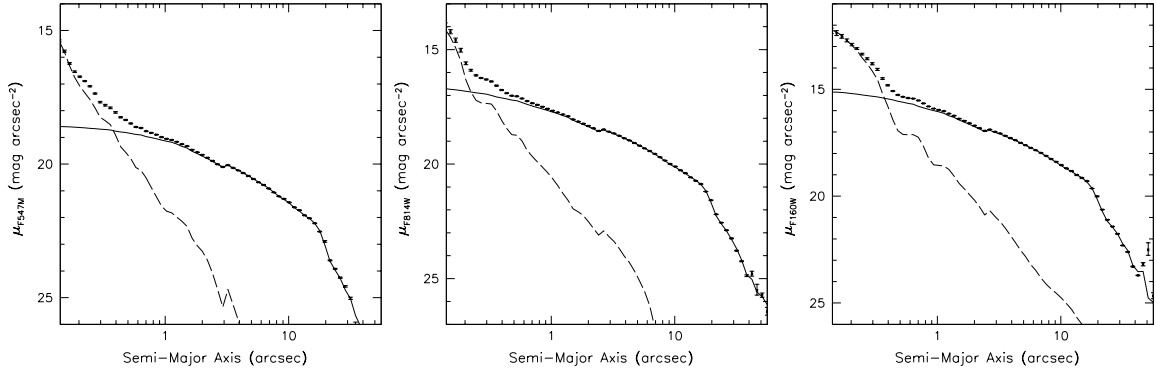}
\caption{One-dimensional surface brightness profiles for UGC\,06728 in F547M, F814W, and F160W ({\it left to right}).  The data points show the surface brightness measurements, while the solid lines show the best-fit galaxy profiles (bulge, bar, ansae, and disk) and the dashed lines show the AGN contribution at the galaxy center. 
}
\label{fig:oned}
\end{figure*}

Fitting was carried out first on the F160W image (see Figure~\ref{fig:galfit}), then the best-fit parameters for that image were used as starting values for the fits to the F547M and F814W images.  Table~\ref{tab:galfit} lists the best-fit parameters for all three images, with magnitudes reported in the vegamag system.  In general, the shapes of the bulge, bar, and disk are found to be very similar across the filters, with similar sizes, \sersic\ indices, and ellipticities.   This is perhaps unsurprising given the fairly red bias of the filters and the lack of star formation in the galaxy.  The disky/boxy parameter always preferred positive values ("boxy" shapes), as would be expected for a strongly barred galaxy like UGC\,06728.  

Using the final best-fit parameters, we created sky-subtracted and AGN-free images of the galaxy in each filter.  In Figure~\ref{fig:oned}, we show one-dimensional surface brightness profiles in each filter, with the galaxy and the AGN displayed separately as the solid and dashed lines respectively, and the total light profile represented by the data points.

The final magnitudes for the entire galaxy and for each of the surface brightness components were converted from the \hst\ filter system to the more commonly-used $V$, $I$, and $H$ filters with synthetic photometry.  Using {\sc synphot}, we  adopted an elliptical galaxy template that was reddened and redshifted to match UGC\,06728, and then measured the magnitude difference between the \hst\ bandpasses and ground-based filters.  We then corrected the $V$, $I$, and $H$-equivalent photometry for Galactic extinction along the line of sight. Based on the \citealt{schlafly11} recalibration of the \citealt{schlegel98} dust map, we adopted  $A_V=0.278$\,mag, $A_I=0.153$\,mag, and $A_H=0.046$\,mag.  In Table~\ref{tab:phot} we provide the galaxy photometry as measured from our image decompositions in the \hst\ filter system, and the extinction-corrected $V$, $I$, and $H$-equivalent photometry.

\begin{deluxetable}{lcccccc}
\input{table.phot.tex}

\tablecomments{HST filter magnitudes are observed magnitudes in the Vegamag system. V, I, and H magnitudes are corrected for Galactic extinction using the  \citet{schlafly11} recalibration of the \citet{schlegel98} dust map: $A_V = 0.278$\,mag, $A_I = 0.1563$\,mag, and $A_H = 0.046$\,mag.}
\end{deluxetable}

\section{Distance}

Based on its redshift and $H_0=74$\,km\,s$^{-1}$\,Mpc$^{-1}$ \citep{riess19}, the estimated distance to UGC\,06728 is 26.4\,Mpc.  However, it is well established that nearby galaxies can suffer from peculiar velocities on the order of 500\,km\,s$^{-1}$ and larger, and so their distances can be significantly different than indicated by their redshifts.

UGC\,06728 shows an excess of point-like sources surrounding the galaxy in all filters.  These are especially apparent in the residuals of the modeled images.  The lack of dust or apparent ongoing star formation indicate that these sources may be globular clusters.  We therefore attempted to constrain the distance to UGC\,06728 through the globular cluster luminosity function (cf.\ the review by \citealt{rejkuba12}).

To begin searching for candidate globular clusters, we followed the standard procedures of identifying and carrying out PSF photometry for all point-like sources in the frames using {\sc DAOPHOT}, {\sc ALLSTAR}, and {\sc ALLFRAME} \citep{stetson87,stetson94}.  Sources were identified separately in each frame, after which the source lists were matched by position.  Only those sources that were identified in all three images were retained, leaving an initial set of 85 sources.

\begin{figure*}
\plottwo{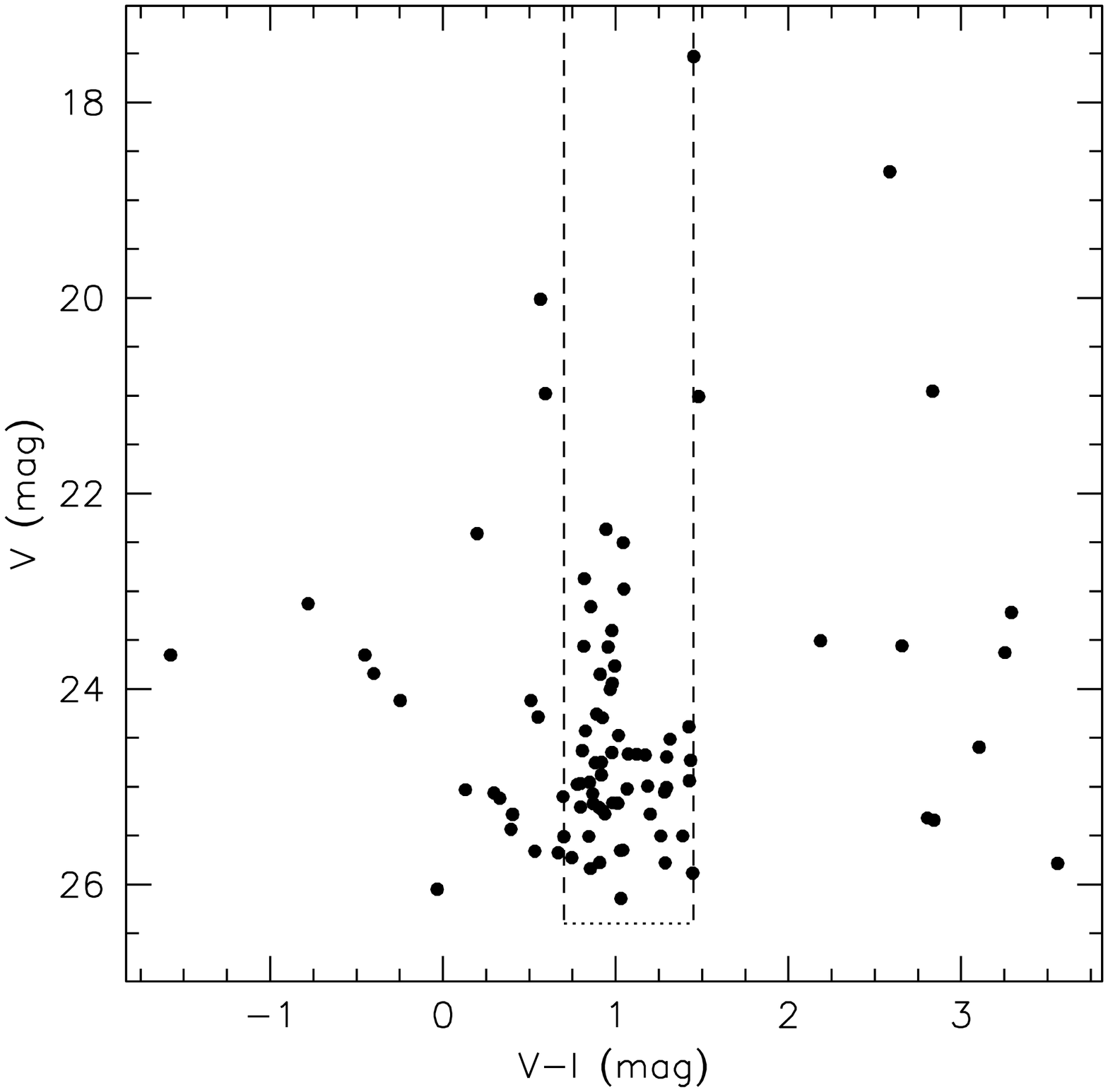}{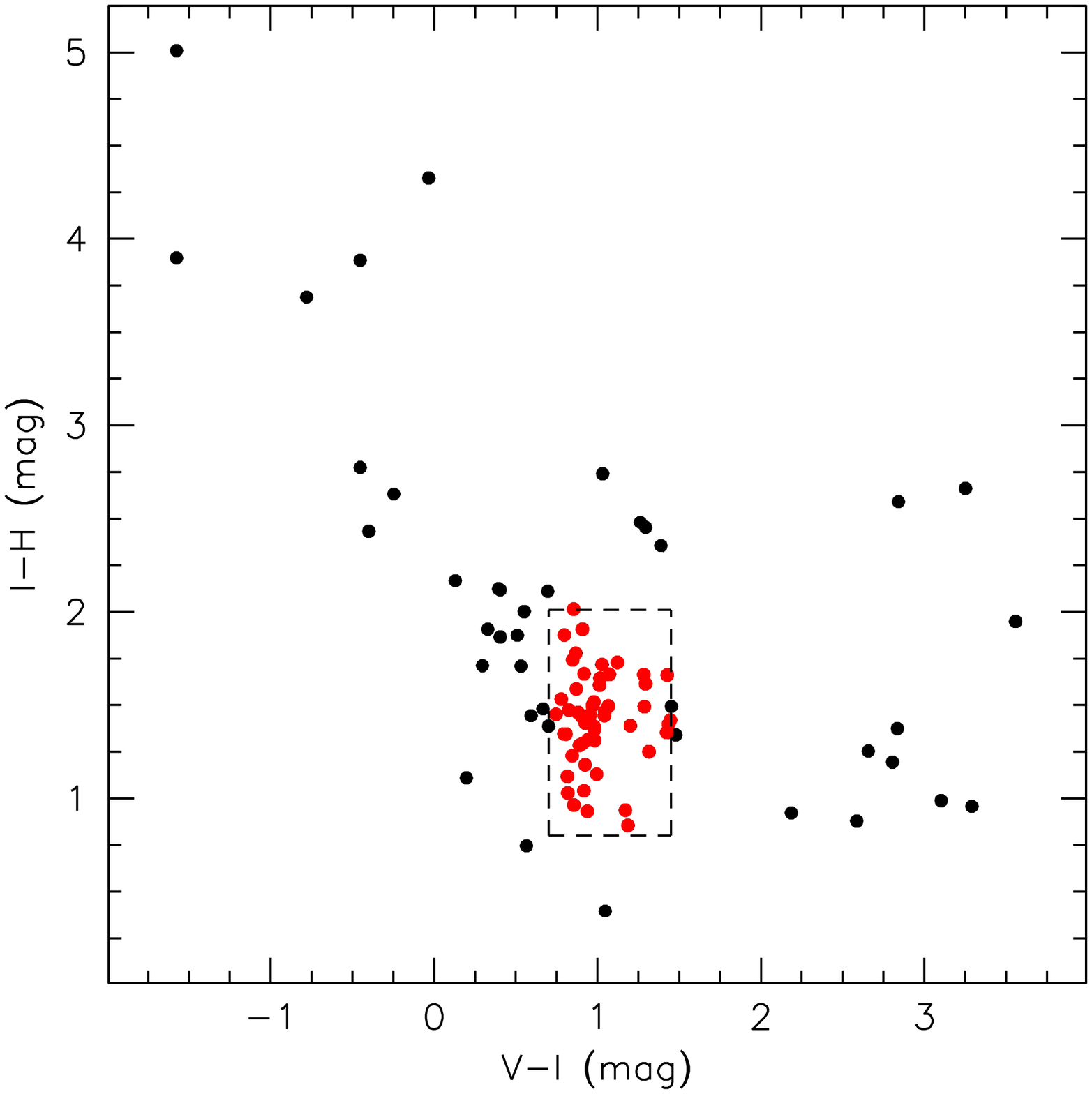}
\caption{ Colors and magnitudes of point-like sources detected in the field of UGC\,06728.  Dashed lines mark the color cuts described in the text.  In the left panel, the dotted horizontal line marks the 50\% completeness level in $V$ as determined from artificial source tests.  In the right panel, the red points inside the dashed rectangle are the final set of 51 globular cluster candidates.}
\label{fig:color}
\end{figure*}

\begin{figure*}
\plotone{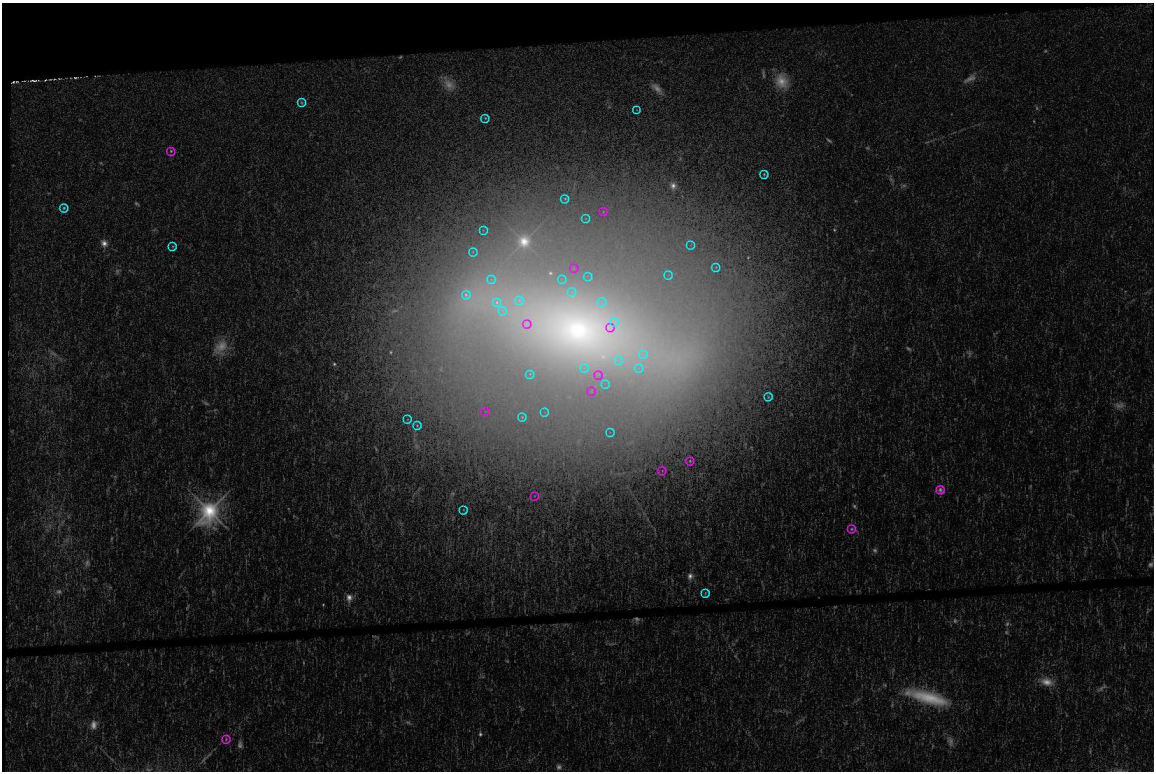}
\caption{F814W image of UGC\,06728 with the locations of the 51 globular cluster candidates marked by cyan circles for the blue population and magenta circles for the red population. 
}
\label{fig:gcs}
\end{figure*}

These 85 sources were then subjected to a color cut, where candidates with extinction-corrected colors of $0.7 < (V-I) < 1.45$\,mag \citep{larsen01} were retained.  These colors correspond to the range of metallicities expected for globular cluster systems, $2.2 < [Fe/H] < 0.2$, and leave us with 55 candidates.  Finally, we examined the $V-H$ and $I-H$ colors of the remaining candidates and discarded a few outliers that were more than 0.5\,mag discrepant with the distributions of colors for the other candidates.  The remaining 51 candidates (Figure~\ref{fig:gcs}) have colors of $0.8 < I-H < 2.0$\,mag and $1.8 < V-H < 3.1$\,mag (see Figure~\ref{fig:color}), which agree with the expectations from stellar population models of globular clusters \citep{cantiello07}.

The systems of globular clusters around galaxies are generally found to consist of two populations, a "red" population and a "blue" population that often appear as a bimodal shape in the $V-I$ colors.  The blue population is often interpreted as the old, metal-poor population while the red population may be either old and metal rich, or may consist of younger globular clusters \citet{rejkuba12}.  While the blue population is found to be ubiquitous and has  similar properties from galaxy to galaxy, the fraction of clusters in the red population decreases for fainter galaxies and their mean color decreases as well \citep{peng06}.  Thus the blue population is preferred for distance estimates based on the globular cluster luminosity function. 

\begin{figure}
\plotone{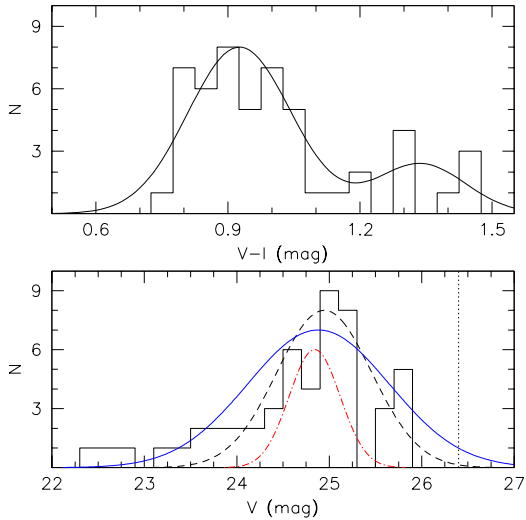}
\caption{{\it Top:} $V-I$ color distribution of globular cluster candidates, with a clear separation between blue and red populations and the best fit double Gaussian overlaid. {\it Bottom:} Distribution of $V-$band magnitudes for globular cluster candidates, with the best fit Gaussians representing the globular cluster luminosity function for the blue population (solid blue line), red population (dot-dash red line), and all 51 candidates (dashed black line). The peak values of the Gaussians are arbitrarily offset for clarity.   The dotted vertical line marks the 50\% completeness limit determined from artificial source tests.
}
\label{fig:gclf}
\end{figure}

The $V-I$ colors of the globular cluster candidates around UGC\,06728 show a clear bimodal signature, indicating that both a blue and red population are present (Figure~\ref{fig:gclf}, top).  Adopting a typical division between blue and red of $V-I=1.05$\,mag, the blue population makes up 73\% of the candidates.  There does not seem to be any difference in the locations of the two populations around the host galaxy, but the small number of red candidates also makes this difficult to accurately determine.

The typical method of characterizing the globular cluster luminosity function is to work with their magnitudes, rather than their luminosities, and to fit the shape of the magnitude distribution with a Gaussian (e.g., \citealt{rejkuba12}).  The distribution is then characterized by the dispersion and by the peak, which is taken to be the turnover magnitude and is the standard candle that can provide an estimate of the galaxy distance.  The assumption of a Gaussian shape is not physically motivated, but instead serves as a useful parameterization that provides a relatively unbiased distance estimate \citep{mclaughlin94}.

We fit the magnitude distribution of the blue population of globular cluster candidates with a Gaussian function as our primary distance indicator.  We also fit the distribution of the red population and the distribution of all 51 candidates as a check on the fit to the blue population.  We find a best-fit turnover magnitude of $V_{\rm TO} = 24.9 \pm 0.1$\,mag for the blue population, and this agrees well with the fits to the red population and to the full set of candidates, with $V_{\rm TO} = 24.8 \pm 0.1$\,mag and $V_{\rm TO} = 24.9 \pm 0.1$\,mag, respectively (Figure~\ref{fig:gclf}, bottom).  

\begin{figure}
\plotone{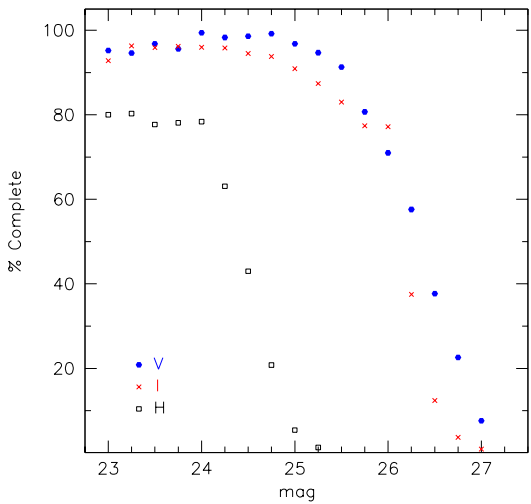}
\caption{ Completeness as a function of magnitude based on artificial source tests conducted in $V$ (filled blue hexagons), $I$ (red crosses), and $H$ (open black squares).} 
\label{fig:ast}
\end{figure}

We assessed the completeness of our globular cluster candidates by running a series of artificial source tests.  In each band, we investigated the completeness as a function of magnitude in steps of 0.25\,mag between $23-27$\,mag.  Into 10 copies of each science frame, we inserted 100 artificial sources using the PSF model created by {\sc DAOPHOT} in our search for globular cluster candidates. Each artificial source was inserted with a random position and with its magnitude allowed to randomly deviate within $\pm0.05$\,mag of the magnitude step.  Photometry was then carried out as before, and the percentage of artificial sources recovered from the frames was recorded. The artificial source tests demonstrate that the globular cluster candidates are more than 50\% complete down to $V=26.4$\,mag (see Figure~\ref{fig:ast}), which is $\sim 1.5$\,mag below $V_{\rm TO}$.  Given the typical colors of $V-I\approx1.0$\,mag and $V-H\approx2.5$\,mag for the candidates, the completeness for candidates at this $V-$band magnitude is higher in $I$ and $H$.  Correcting for completeness does not change the best-fit turnover magnitudes except for a slight shift when the full sample is used, from $V_{\rm TO} = 24.9 \pm 0.1$\,mag to $V_{\rm TO} = 25.0 \pm 0.1$\,mag.  Based on the fits to the magnitude distribution and the artificial source tests, we adopt $V_{\rm TO} = 24.9 \pm 0.2$\,mag for the apparent $V-$band turnover magnitude of the globular cluster candidates around UGC\,06728.

The calibration of the $V-$band turnover magnitude for blue population clusters in the Milky Way, M31, and 17 other nearby galaxies has been found to be $-7.66 \pm 0.09$\,mag \citep{rejkuba12}.  This absolute turnover magnitude is calibrated to an assumed LMC distance modulus of $18.5$\,mag that is supported by the recent determination of a distance modulus to the LMC of  $18.477 \pm 0.026$\,mag using detached eclipsing binary systems \citep{pietrzynski19}.  Adopting this calibration, we determine a distance modulus to UGC\,06728 of $m-M=32.56\pm0.22$, which implies a distance of $D=32.5\pm3.5$\,Mpc.

\section{Discussion}

Based on the globular cluster candidates identified around UGC\,06728, the distance to the galaxy is somewhat larger than predicted by its optical redshift of $z=0.0065$ \citep{falco99}.  Depending on the adopted value of $H_0$, this implies that peculiar velocities of $-240$\,km\,s$^{-1}$ ($H_0=67.4$\,km\,s$^{-1}$\,Mpc$^{-1}$; \citealt{planck18}) to $-450$\,km\,s$^{-1}$ ($H_0=74$\,km\,s$^{-1}$\,Mpc$^{-1}$; \citealt{riess19}) are affecting the galaxy, and there does appear to be a loose grouping of galaxies in the vicinity of UGC\,06728 that could be responsible.  With a larger distance, all prior luminosity estimates that relied on the redshift of the galaxy are underpredicted by $25-50$\%.

\subsection{AGN Luminosity}

For nearby galaxies like UGC\,06728, one of the complications in reverberation  mapping programs is that the large spectroscopic slit employed to minimize seeing effects in the observations also results in significant starlight dilution of the AGN continuum (e.g., \citealt{bentz09b,bentz13}).  Based on low-resolution seeing-limited images, \citet{bentz16b} estimated the starlight contribution through the 5\arcsec$\times$5\arcsec slit (PA=0\degr) to be $f_{gal}=1.09\pm0.22\times10^{15}$\,\flam\ at $5100\times(1+z)$\,\AA.

With the AGN-subtracted F547M image we can improve on this estimate.   Using a rectangular aperture of the same angular size and rotation as the spectroscopic monitoring aperture, we first measure the counts in the AGN- and sky-subtracted F547M image.  The exposure time and inverse sensitivity ({\tt photflam} keyword) allow us to recover the flux density from the counts.  We then applied a small color correction to account for the slight difference between the effective center of the F547M filter and $5100\times(1+z)$\,\AA.  We find $f_{gal}=1.11\pm0.11\times10^{15}$\,\flam\ at $5100\times(1+z)$\,\AA, which is remarkably consistent with the ground-based estimate.  The observed spectroscopic flux density through the ground-based monitoring aperture was measured by \citet{bentz16b} to be $f = (2.21\pm0.20) \times 10^{15}$\,\flam\, and so the starlight contribution is 50\% of the total flux density at $5100\times(1+z)$\,\AA.  The AGN flux density is then $f_{\rm AGN}= (1.10\pm0.12) \times 10^{-15}$\,\flam.

Adopting the distance implied by the globular cluster luminosity function gives an AGN luminosity of $\log \lambda L_{5100} = 41.97 \pm 0.11$\,erg\,s$^{-1}$. This is about 40\% larger than estimated by \citet{bentz16b}, although it is consistent with the previous estimate given the generous uncertainty adopted for the distance in that calculation. Thus, when the AGN luminosity is combined with the broad H$\beta$ time delay, UGC\,06728 is still found to be consistent with the \rl\ relationship of \citet{bentz13}.

\subsection{Galaxy Luminosity and Globular Clusters}

With the $V$, $I$, and $H$-equivalent photometry and the distance predicted by the globular cluster luminosity function, we can constrain the luminosity of the galaxy.  We first determined the $k$-corrections needed to transform the galaxy photometry into $z=0$-equivalent photometry.  The small redshift of $z=0.00652$ meant that the $k$-corrections were quite small, $< 0.02$\,mag in all three filters.  The $V-$band absolute magnitude of the galaxy is thus $M_V=-18.77\pm0.31$\,mag

The globular cluster luminosity function has been found to be reliable as a distance indicator for massive galaxies down to at least $M_B \approx -18$ \citep{rejkuba12}.  Studies of dwarf galaxies in the Virgo and Fornax clusters find a systematic offset in the $V-$band turnover magnitude of their globular cluster systems \citep{villegas10}, suggesting that caution must be taken when using globular clusters around dwarf galaxies to estimate distances.  However, a large study of local dwarf galaxies in loose groups by \citet{georgiev09} found no such bias, indicating that perhaps environment plays a role in the reliability of globular clusters as distance indicators for dwarf galaxies. 

While we do not have $B-$band photometry of UGC\,06728, we can estimate it by adopting the typical color of an S0 galaxy,  $B-V\approx0.87$\,mag \citep{roberts94}.  Combined with the $V-$band equivalent photometry, the $B-$band absolute magnitude is $M_B \approx -17.9$\,mag.  This places UGC\,06728 at the low end of the galaxy brightness range where the globular cluster luminosity function may be considered reliable.  Furthermore, the location of UGC\,06728 within a loose group of galaxies, rather than a dense cluster, lends additional support to the reliability of the distance predicted by its globular clusters. Finally, the dispersion of our Gaussian fit to the distribution of globular cluster candidate magnitudes, $\sigma\approx0.8$\,mag for the blue population, is slightly lower than but within the scatter around the relationship between galaxy $M_B$ and globular cluster luminosity function dispersion \citep{jordan07}.

\subsection{Galaxy Stellar Mass}

The stellar $M/L$ of galaxies is often predicted by the use of two-filter colors.  Comparison of many different prescriptions for estimating $M/L$ by \citet{roediger15} finds that prescriptions based on multiple filters generally predict more reliable values.  Unfortunately, such studies have generally focused on SDSS filter colors and not the Johnson-Cousins set.  \citet{zhang17} analyzed the effects of metallicity, age, dust, and other details on the predicted stellar masses of local universe galaxies and find that $V$ is the optimal luminance band for estimating $M/L$ from single colors.

Using the prescriptions of \citet{bell01} for $V-H=2.82$\,mag predicts $M/L (V)=3.13 M_{\odot}/L_{\odot}$. The $V-I$ color, however, predicts a slightly higher value of $M/L (V)=4.13 M_{\odot}/L_{\odot}$.  Examining the full combination of  colors and passbands predicts a range of stellar masses covering $\log M_{\star}/M_{\odot}=9.5-10$, but with the median value being $\log M_{\star}/M_{\odot}=9.9\pm0.2$.  If we instead adopt the prescriptions of \citet{zhang17} based on the FSPS stellar population models, we find nearly the same result with a median value of $\log M_{\star}/M_{\odot}=9.9\pm0.3$ based on $M/L(V)$ and $V-I$. 

This is quite similar to the estimate of
\citet{bentz16b} based on seeing-limited images, although a different set of prescriptions were used to predict the stellar $M/L$ from the $g-r$ color, and a smaller distance of $D_L=27$\,Mpc was assumed.

With a central black hole mass of $M_{\rm BH}=7\times10^5$\,\msun, UGC\,06728 appears to follow the trend of lower $M_{\rm BH}/M_{\star}$ ratios for lower mass black holes \citep{bentz18}, with $M_{\rm BH}/M_{\star}\approx 10^{-4}$.

\subsection{Implications}

The presence of strong ansae at the ends of the bar in UGC\,06728 is not unusual, as $\sim40$\% of SB0 galaxies have been found to exhibit ansae, though they are rarer for later galaxy types \citep{martinez07}.  The large bar/total fraction, ranging from $0.23-0.27$ across the three filters and reaching almost 0.4 in F814W if the ansae are included, does appear to be somewhat unusual, however. A study of 300 galaxies with $i-$band imaging from SDSS found a median bar/total fraction of $0.09\pm0.05$ \citep{gadotti09}, with very few bar/total fractions greater than 0.2.  However, UGC\,06728 is at the limit of the range of galaxy properties included in that study.  Given the limitations of the SDSS imaging, galaxies with $\log M_{\star}/M_{\odot}<10$ were excluded and the study was biased against the detection of bars $\lesssim 2$\,kpc in length.  Even though the bar spans most of the galaxy, the compact nature of UGC\,06728 means the total bar length is only $\sim 1.7$\,kpc from end to end.  

With the non-detection of \ion{H}{1} emission from UGC\,06728  \citep{robinson19} and the smoothness of the galaxy in all three filters showing no ongoing star formation, there appears to be a low gas content in the galaxy.  This is also supported by the lack of apparent dust in the images and the clear view to the nucleus reported by \citet{walton13}, even though the disk axis ratio suggests that we are viewing the galaxy at a moderate inclination ($\sim 40^{\circ}$).

The outer regions of the galaxy disk are symmetric in shape down to the lowest levels visible in the \hst\ imaging, showing no signature of recent mergers, neither major nor minor.   Even when stacking all of the images together, there are no apparent tidal features or shells.  Thus, the strong nuclear activity and high spin of the black hole ($a>0.7$) are somewhat mysterious given that there is no obvious fuel supply.  If bar color is an indication of bar age, as has been suggested by some authors (e.g., \citealt{gadotti06}), then the red bar color of $V-I=1.4$\,mag may suggest that the bar in UGC\,06728 is quite evolved.   Funneling of gas to the nucleus via the bar could have happened long ago, and UGC\,06728 may be nearing the end of its capabilities for sustained, strong nuclear activity.   

\begin{deluxetable}{lrcll}
\input{table.derived.tex}
\end{deluxetable}

\section{Summary}

The dwarf Seyfert UGC\,06728 resides in a barred lenticular galaxy with prominent ansae at the ends of the bar.  Using high-resolution \hst\ imaging through three optical and near-infrared filters, we have derived several properties of the galaxy, which are summarized in Table~\ref{tab:derived}.  Two-dimensional image decompositions were carried out on all three images, allowing the AGN to be cleanly separated from the galaxy and detailed surface brightness profiles be constrained in each filter.  We have identified globular cluster candidates and employed the globular cluster luminosity function method to estimate a distance to the galaxy of $D=32.5\pm3.5$\,Mpc.  Finally, we have derived the starlight-corrected luminosity of the AGN, the absolute magnitude of the galaxy, and the stellar mass of the galaxy.  As the least massive Seyfert with a direct black hole mass constraint and a spin constraint, the detailed analysis of the host galaxy properties presented here can be expected to help inform black hole evolutionary models.

\acknowledgements

We thank the referee for comments that improved the presentation of this work.  We thank Judy Schmidt for taking the time to create a beautiful composite image of UGC\,06728 based on the data presented in this manuscript.
MCB gratefully acknowledges support of this work through grant HST GO-15263 from the Space Telescope Science Institute, which is operated by the Association of Universities for Research in Astronomy, Inc., under NASA contract NAS5-26555. This research has made use of the NASA/IPAC Extragalactic Database (NED) which is operated by the Jet Propulsion Laboratory, California Institute of Technology, under contract with the National Aeronautics and Space Administration and the SIMBAD database, operated at CDS, Strasbourg, France.

\bibliographystyle{apj}

\end{document}

%% file: table.obs.tex
\tablecolumns{4}
\tablewidth{0pt}
\tablecaption{Observations}
\tablehead{
\colhead{Filter} &
\colhead{UT$_{\rm mid}$} &
\colhead{Exp.\ Time} &
\colhead{Dataset} 
\\
\colhead{} &
\colhead{(yyyy-mm-dd hh:mm:ss)} &
\colhead{($s$)} &
\colhead{} 
}
\startdata
F547M  & 2018-03-16 12:37:22 &   1770.0  & IDGO01010, 01020   \\
F814W  & 2018-03-16 14:28:36 &   1950.0  & IDGO01040, 01050   \\
F160W  & 2018-03-16 13:32:57 &   1596.9  & IDGO01030   
\label{tab:obs}
\enddata

%% file: table.galfit.tex
\tablecolumns{12}
\tablewidth{0pt}
\tablecaption{Image Decompositions}
\tablehead{
\colhead{No.} &
\colhead{PSF+sky} &
\colhead{$\Delta x$ ($\arcsec$)} &
\colhead{$\Delta y$ ($\arcsec$)} &
\colhead{$m_{\rm Vega}$ (mag)} &
\colhead{\nodata} &
\colhead{Sky (cts)} &
\colhead{$d$sky/$d$x ($10^{-4}$ cts)} & 
\colhead{$d$sky/$d$y ($10^{-4}$ cts)} &
\colhead{C0} &
\colhead{$f/f_{\rm tot}$} &
\colhead{Note} \\
\colhead{} &
\colhead{sersic} &
\colhead{$\Delta x$ ($\arcsec$)} &
\colhead{$\Delta y$ ($\arcsec$)} &
\colhead{$m_{\rm Vega}$ (mag)} &
\colhead{$r_e$ ($\arcsec$)} &
\colhead{$n$} &
\colhead{$b/a$} & 
\colhead{P.A.\ ($\degr$)} &
\colhead{} &
\colhead{} &
\colhead{} \\
\colhead{} &
\colhead{sersic3} &
\colhead{$\Delta x$ ($\arcsec$)} &
\colhead{$\Delta y$ ($\arcsec$)} &
\colhead{$\Sigma_{b,Vega}$ (mag)} &
\colhead{$r_e$ ($\arcsec$)} &
\colhead{$n$} &
\colhead{$b/a$} & 
\colhead{P.A.\ ($\degr$)} &
\colhead{} &
\colhead{} &
\colhead{} \\
\colhead{} &
\colhead{radial} &
\colhead{$\Delta x$ ($\arcsec$)} &
\colhead{$\Delta y$ ($\arcsec$)} &
\colhead{\nodata} &
\colhead{$r_{break}$ ($\arcsec$)} &
\colhead{$r_{soft}$ ($\arcsec$)} &
\colhead{$b/a$} & 
\colhead{P.A.\ ($\degr$)} &
\colhead{} &
\colhead{} &
\colhead{} 
}
\startdata

\multicolumn{11}{c}{F547M}
\\ \hline \\

1,2	& PSF+sky &	0	        & 0	     & 16.20	& \nodata	& 17.0	& -0.19	& 2.68	&&&	AGN, sky \\
3	& sersic  &	0.07	    & -0.09	 & 16.55	& 2.11	    & 1.0	& 0.72	& 75.3	& 0.10 & 0.11 &bulge \\
4	& sersic  &	0.08	    & -0.13	 & 15.72	& 6.82	    & 0.5	& 0.45	& 77.0	& 0.19 & 0.23 & bar \\
5	& sersic3 &	-7.32	    & 2.39	 & 22.12	& 227.38	& [1.0]	& 0.03	& 73.1	&&	0.09 & ansae \\
	& radial, inner	& 4.66	& 3.95	 & \nodata  & 45.05	    & 42.33	& 0.40	& -51.5 &&&	\\
	& radial, outer	& -9.31	& -17.10 & \nodata  & 66.86	    & 81.83	& 0.15	& -28.8	&&&	\\
6	& sersic  &	0.04	    & 0.06	 & 14.72	& 14.85	    & [1.0]	& 0.76	& -84.5	&&	0.57 & disk \\

\\ \hline
\multicolumn{11}{c}{F814W}
\\ \hline \\

1,2 & PSF+sky &	0	        & 0	    & 15.34	  & \nodata	& 30.4	& -5.00	& 2.72	&&&	AGN, sky \\
3	& sersic  &	0.03	    & -0.03	& 15.44	  & 1.60	& 1.5	& 0.82	& 75.0	& 0.04 & 0.08 & bulge \\
4	& sersic  &	0.09	    & -0.15	& 14.16	  & 5.99	& 0.7	& 0.49	& 77.8	& 0.26 & 0.27 & bar \\
5	& sersic3 &	-8.49	    & 2.77	& 21.04	  & 220.54	& [1.0]	& 0.03	& 72.8	&&	0.10 & ansae \\
	& radial, inner & 6.77	& 6.57	& \nodata &	73.49	& 78.47	& 0.26	& -46.3	&&&	\\
	& radial, outer & -3.48	& -6.25	& \nodata &	60.25	& 73.10	& 0.17	& -29.7	&&&	\\
6	& sersic  &	-0.08	    & 0.12	& 13.39	  & 15.96	& [1.0]	& 0.80	& -82.4	&& 0.55 & disk \\

\\ \hline
\multicolumn{11}{c}{F160W}
\\ \hline \\
1,2 & PSF+sky &	    0	     & 0	  & 13.94	& \nodata & 1.9   & -5.93 & -8.07  &&&	AGN, sky \\
3	& sersic  &	    0.06	 & 0.03	  & 13.45	& 1.80	  & 1.5	  & 0.76  & 80.2   & 0.25 &	0.13 &bulge \\
4	& sersic  &	    0.04	 & -0.03  & 12.64	& 6.748	  & 0.6	  & 0.43  & 76.2   & 0.30 & 0.26 &	bar \\
5	& sersic3 &	    -8.93	 & 2.87   & 19.16   & 221.66  & [1.0] & 0.03  & 73.2   && 0.09 &	ansae \\
	& radial, inner & 0.73	 & 0.43	  & \nodata & 36.93	  & 35.39 & 0.49  & -55.1  &&&	\\
	& radial, outer & -2.75  & -5.06  & \nodata & 35.69	  & 40.97 & 0.29  & -29.5  &&&	\\
6	& sersic  &	    0.02	 &-0.05	  & 11.91	& 15.75	  & [1.0] & 0.78  & -82.2  && 0.52 &	disk \\

\label{tab:galfit}
\enddata

%% file: table.phot.tex
\renewcommand{\arraystretch}{2}
\tablecolumns{7}
\tablewidth{0pt}
\tablecaption{Galaxy Photometry}
\tablehead{
\colhead{Component} &
\colhead{$m_{F547M}$} &
\colhead{$m_{F814W}$} &
\colhead{$m_{F160W}$} &
\colhead{V} &
\colhead{V-I} &
\colhead{V-H} \\
\colhead{} &
\colhead{(mag)} &
\colhead{(mag)} &
\colhead{(mag)} &
\colhead{(mag)} &
\colhead{(mag)} &
\colhead{(mag)} 
}
\startdata
Bulge &	16.55 &	15.44 &	13.45 &	16.24 &	0.92 &	3.00 \\
Bar	  & 15.72 &	14.16 &	12.64 &	15.42 &	1.38 &	2.98 \\
Ansae &	16.68 &	15.21 &	13.79 &	16.37 &	1.29 &	2.79 \\
Disk  &	14.72 &	13.39 &	11.91 &	14.42 &	1.15 &	2.71 \\
Total &	14.11 &	12.73 &	11.20 &	13.81 &	1.20 &	2.82 
\label{tab:phot}
\enddata

%% file: table.derived.tex
\renewcommand{\arraystretch}{2}
\tablecolumns{5}
\tablewidth{0pt}
\tablecaption{Derived Quantities}
\tablehead{
\colhead{Quantity} &
\multicolumn{3}{c}{Value} &
\colhead{Units}
}
\startdata
$m-M$             &   32.56  & $\pm$ & 0.22  &  mag \\
D	              &   32.5   & $\pm$ & 3.5   &  Mpc \\
$\log \lambda L_{5100}$\,(AGN) & 41.97 & $\pm$ & 0.11 & erg\,s$^{-1}$ \\
$M_V$\,(gal)      &   -18.77 & $\pm$ & 0.31  &  mag \\
$\log M_{\star}/M_{\odot}$  &   9.9   & $\pm$ & 0.2   & 	 
\label{tab:derived}
\enddata 